\documentclass[12pt]{article}
\usepackage{epsfig}
\usepackage{amsmath}
\usepackage{cite}
\usepackage{amssymb}
\newlength{\defbaselineskip}
\newcommand{\setlinespacing}[1]%
           {\setlength{\baselineskip}{#1 \defbaselineskip}}

\begin{document}

\title{Effect of nuclear matter incompressibility on the \textsuperscript{16}O+\textsuperscript{208}Pb system}

\author{O.N.Ghodsi\thanks{Email: o.nghodsi@umz.ac.ir}, F.Torabi\\
\\
{\small {\em  Department of Physics, Faculty of Science, University of Mazandaran}}\\
{\small {\em P.O.Box 47415-416, Babolsar, Iran}}\\
}
\date{}
\maketitle

\begin{abstract}
\noindent To analyze the property of nuclear matter in the \textsuperscript{16}O+\textsuperscript{208}Pb collision system,
the internuclear potential of the fusion reaction is calculated by using the Skyrme forces associated with an
extensive nuclear matter incompressibility $K$ range in the semiclassical energy density formalism. Comparison
of the experimental fusion cross sections and those obtained by using potentials derived from different forces with various $K$ values shows that the incompressibility of nuclear matter changes during the fusion process at different bombarding energies. The results indicate that, as the energy increases, the nuclear matter becomes more incompressible.
\\
\\
\\
\\
\\

\end{abstract}

\newpage
\setlinespacing{1.5}
\noindent{\bf{I. Introduction}\\}

Fusion in heavy-ion reactions has been one of the most extensively studied topics in nuclear physics over the last decades \cite{1,2,3,4,5,6,7,8,9,10,11,12,13,14}. Various attempts have been made to explain this
phenomenon by using a variety of theoretical models based on different assumptions. Taking into
account the dynamical mechanism of the fusion process, the interaction
potential between two nuclei can be determined by using dynamic approaches such as quantum molecular dynamics and time-dependent Hartree\textendash
Fock theory \cite{15,16,17,18,19,20}. According to the frozen-density approximation, fusion
reactions can be also analyzed by using static approaches such as the
double-folding model and energy-density formalism
\cite{21,22,23,24,25,26}. By employing different effective
nucleon-nucleon interactions in these models and methods, a large number of heavy-ion fusion reactions have been investigated in theoretical
low-energy nuclear physics. Among them, the
\textsuperscript{16}O+\textsuperscript{208}Pb system is a candidate
that has been widely studied by using static and dynamic approaches
\cite{27,28,29,30}. Some studies have shown that analysis of the fusion cross-section data of this heavy-ion reaction
can help understand the importance of different factors in calculations of the interaction potential, including the energy dependence of the
barrier \cite{27} and the incompressibility of nuclear matter
\cite{30}.

Nuclear matter incompressibility ($K$) is a key component of the
nuclear matter equation of state (EOS) and has been one of the
interesting subjects in studies of heavy-ion fusion reactions. Different versions of the effective
interactions resulting in different $K$ values have been used to
investigate the role of nuclear matter incompressibility in
heavy-ion fusion processes \cite{30,31,32}. The results obtained
revealed that theoretical fusion data is sensitive to the value of $K$. Therefore,
describing the heavy-ion reaction by using different effective
interactions with different $K$ values may allow exploration of variations in the incompressibility of nuclear matter during the fusion process at different bombarding energies.

Accordingly, in the present study, we are motivated to
examine this variation within the
\textsuperscript{16}O+\textsuperscript{208}Pb system. For this purpose,
the interaction potential of the chosen system was calculated by using
different Skyrme forces associated with $K$ values ranging from 234 to 370 MeV in the semiclassical energy-density
formalism. With respect to each force, the neutron and proton densities obtained
by the self-consistent quantum-mechanical Hartree\textendash Fock\textendash Bogoliubov (HFB) method were also employed in this formalism. Based on the best agreement achieved between the theoretical fusion
cross sections obtained by the potentials derived from
different forces and the experimental data, we have shown variation in
the nuclear matter incompressibility within the \textsuperscript{16}O+\textsuperscript{208}Pb system at different
bombarding energies.

This paper is organized as follows: Section II introduces the Skyrme
energy-density-functional model and describes the properties of the
colliding nuclei based on the effective interactions employed in
this model. Section III presents the calculations and results of analysis of the \textsuperscript{16}O+\textsuperscript{208}Pb system by
using different forces yielding various incompressibility values.
Finally, Sec. IV draws the conclusions of this paper.
\\

\noindent{\bf {II. Theoretical Formalism}}\\

\noindent{\bf\small{A. Semiclassical expression of the Skyrme energy-density functional}}

In the energy-density-functional model, the nuclear potential between the interacting nuclei, as a function of separation distance $R$, is given by

\begin{equation} \label{1}
V_{N}(R)=E_{T}(R)-(E_{1}+E_{2}),
\end{equation}
\begin{equation} \label{2}
E_{T}(R)=\int{\mathcal{E}\left[ \rho_{1p}(\vec{r})+\rho_{2p}(\vec{r}-\vec{R}),\rho_{1n}(\vec{r})+\rho_{2n}(\vec{r}-\vec{R})\right] \mathrm{d^3}r,}
\end{equation}
\begin{equation} \label{3}
E_{1}=\int{\mathcal{E}\left[ \rho_{1p}(\vec{r}),\rho_{1n}(\vec{r})\right] \mathrm{d^3}r,}
\end{equation}
\begin{equation} \label{4}
E_{2}=\int{\mathcal{E}\left[ \rho_{2p}(\vec{r}),\rho_{2n}(\vec{r})\right] \mathrm{d^3}r,}
\end{equation}
\\
where $E_1$ and $E_2$ denote the energy of the noninteracting
nuclei and $E_{T}(R)$ expresses the energy of the composite system.
In these equations, the Skyrme energy density $\mathcal{E}(\vec{r})$
is defined as

\begin{equation}
\begin{aligned}
\mathcal{E}(\vec{r})= & \frac{\hbar^{2}}{2m}\tau+\frac{1}{2}t_{0}\left[\left( 1+\frac{1}{2}x_{0}\right) \rho^{2}-\left(x_{0}+\frac{1}{2}\right) ({\rho_{n}}^2+{\rho_{p}}^2)\right] \\
 &+\frac{1}{12}t_{3}\rho^{\alpha}\left[\left( 1+\frac{1}{2}x_{3}\right) \rho^{2}-\left( x_{3}+\frac{1}{2}\right) ({\rho_{n}}^2+{\rho_{p}}^2)\right] \\
 &+\frac{1}{4}\left[t_{1}\left( 1+\frac{1}{2}x_{1}\right) +t_{2}\left(1+\frac{1}{2}x_{2}\right) \right] (\rho\tau)\\
 &-\dfrac{1}{4}\left[t_{1}\left(x_{1}+\dfrac{1}{2}\right) -t_{2}\left( x_{2}+\frac{1}{2}\right) \right] (\rho_{n}\tau_{n}+\rho_{p}\tau_{p}) \\
 &+\frac{1}{16}\left[3t_{1}\left(1+\frac{1}{2}x_{1}\right) -t_{2}\left(1+\frac{1}{2}x_{2}\right) \right](\vec{\nabla}\rho)^{2}\\
 &-\frac{1}{16}\left[3t_{1}\left(x_{1}+\frac{1}{2}\right) +t_{2}\left(x_{2}+\frac{1}{2}\right) \right] \left[(\vec{\nabla}\rho_{n})^{2}+(\vec{\nabla}\rho_{p})^{2}\right]\\
 &+\frac{1}{2}W_{0}\left[\vec{J}.\vec{\nabla}\rho+\vec{J_{n}}.\vec{\nabla}\rho_{n}+\vec{J_{p}}.\vec{\nabla}\rho_{p}\right].
 \end{aligned}
\end{equation}
\

Here, $t_0$, $t_1$, $t_2$, $t_3$, $x_0$, $x_1$, $x_2$, $x_3$,
$\alpha$ and $W_0$ are the Skyrme force parameters determined by
fitting different properties of nuclei. $m$ is the nucleon mass, and
$\rho=\rho_{n}+\rho_{p}$, $\tau=\tau_{n}+\tau_{p}$, and
$\vec{J}=\vec{J_{n}}+\vec{J_{p}}$ are the nuclear, kinetic, and
spin-orbit densities, respectively. The kinetic energy and
spin-orbit densities are estimated in the semiclassical
extended Thomas\textendash Fermi model (ETF).

Taking into consideration the $ \hbar^2 $ correction terms in this model, the functional form of the kinetic-energy density is given by $(q=n$ or $ p)$,
\begin{equation}\label{6}
\begin{aligned}
{\tau}_{q}(\vec{r}){}=&\frac{3}{5}(3\pi^{2})^{\frac{2}{3}}{\rho_{q}}^{\frac{5}{3}}+\dfrac{1}{36}\frac{({\vec{\nabla}{\rho_{q}})}^{2}}{\rho_{q}}+\dfrac{1}{3}\Delta\rho_{q}+\frac{1}{6}\frac{\vec{\nabla}\rho_{q}.\vec{\nabla}f_{q}}{f_{q}}+\dfrac{1}{6}\rho_{q}\frac{\Delta{f_{q}}}{f_{q}}-\frac{1}{12}\rho_{q}\left( \frac{\vec{\nabla}f_{q}}{f_{q}}\right) ^{2}\\
&+\frac{1}{2}\rho_{q}\left( \frac{2m}{\hbar^{2}}\right)^{2}\left( \dfrac{W_{0}}{2}\frac{\vec{\nabla}{(\rho+\rho_{q})}}{f_{q}}\right) ^{2},
\end{aligned}
\end{equation}
where the effective mass form factor $f_{q}(\vec{r})$ takes the following form:
\\
\begin{equation} \label{7}
f_{q}(\vec{r})=1+\dfrac{2m}{\hbar^{2}}\frac{1}{4}\left[t_{1}\left(1+\frac{x_{1}}{2}\right) +t_{2}\left( 1+\frac{x_{2}}{2}\right) \right] \rho(\vec{r})-\dfrac{2m}{\hbar^{2}}\frac{1}{4}\left[t_{1}\left( x_{1}+\frac{1}{2}\right) -t_{2}\left( x_{2}+\frac{1}{2}\right) \right]\rho_{q}(\vec{r}).
\end{equation}

Because spin is intrinsically a quantum-mechanical
property with no direct classical counterpart, the expression of $
(\vec{J} ) $ in the ETF model is 

\begin{equation} \label{8}
{\vec{J}}_{q}(\vec{r})=-\dfrac{2m}{\hbar^{2}}\dfrac{1}{2}W_{0}\frac{1}{f_{q}}\rho_{q}\vec{\nabla}(\rho+\rho_{q}).
\end{equation}

By using these equations, the nuclear part of the interaction
potential, $V_N (R)$, is determined by knowledge of the density
distributions of the projectile and target nuclei. Then, assuming that
$\rho^{(i)}_{ch}\approx e\rho^{(i)}_{p}$, the Coulomb part is added
to the calculations as
\\
\begin{equation} \label{9}
V_{C}(R)=\int\dfrac{{\rho}^{(1)}_{ch}(\vec{r_{1}}){\rho}^{(2)}_{ch}(\vec{r_{2}})}{\vert\vec{R}+\vec{r_{2}}-\vec{r_{1}}\vert}\mathrm{d^3}{r_1}\mathrm{d^3}{r_2}.
\end{equation}
\\
\noindent{\bf\small{B. Properties of the interacting nuclei}}

To date, numerous parametrizations of the Skyrme effective
interaction have been published and many of them have been applied
in mean-field theories for a variety of purposes. In the present study,
some of the available effective interactions that result in an
EOS with an extensive range of $K$ values are employed to study the nuclear
matter incompressibility in the
\textsuperscript{16}O+\textsuperscript{208}Pb system. The selected forces are SkSC4 \cite{33}, Es \cite{34}, SKXce
\cite{35}, E \cite{34}, and SI \cite{36} with the incompressibility
range between 234 and 370 MeV. Based on each force, the
neutron and proton densities of the \textsuperscript{16}O and
\textsuperscript{208}Pb nuclei were computed by using the microscopic
HFB method because many properties of the finite nuclei
can be described by this approximation. For instance, Fig. 1 shows the radial
density distributions obtained from these calculations based on the SkSC4 and SI parameter sets.

By using the density distributions calculated in the HFB approach, it was
found that all the selected Skyrme forces can reproduce the
experimental binding energies and root-mean-square charge radii of
the chosen nuclei with the relative deviations less than $4.69\%$
and $2.88\%$, respectively. Figure 2 shows the percentage of relative deviations of
the theoretical binding energies and root-mean-square charge radii
from their corresponding experimental data for the SkSC4, Es, SKXce, E, and SI Skyrme forces. These effective
forces, which can describe the ground-state properties of the \textsuperscript{16}O and \textsuperscript{208}Pb nuclei with
reasonable accuracy, are applied to evaluate the nucleus-nucleus
potential in the described energy-density-functional model.
 \\

\
\noindent{\bf{III. Calculations and Results}\\}

To perform the calculations in the energy-density
formalism, based on each of the selected Skyrme forces, the two-parameter Fermi density distributions were determined by using the parameters obtained from fitting the results of HFB calculations. The
calculated diffuseness parameters of the neutron- and proton-density
distributions for the \textsuperscript{16}O and \textsuperscript{208}Pb
nuclei are illustrated in Fig. 3 by using the SkSC4, Es, SKXce, E, and SI
Skyrme forces. Employing the determined densities, together with their
corresponding Skyrme interactions, we evaluated the interaction
potential of the \textsuperscript{16}O+\textsuperscript{208}Pb system.
The characteristics of the calculated fusion barriers, i.e., barrier
height and position, are displayed in Fig. 4 based on the Skyrme forces. The results clearly show that increasing the
value of $K$, increases the fusion barrier height and decreases the value of the barrier position. Also,
from Figs. 3 and 4, and due to the fact that surface nucleons play
a significant role in heavy-ion reactions, one can find that the use of smaller diffuseness parameters in the density distributions decreases the attraction energy and consequently increases the barrier height. 

By using the nucleus-nucleus potentials derived from different Skyrme
forces we analyze here the fusion cross sections of the
\textsuperscript{16}O+\textsuperscript{208}Pb system in different
energy ranges, i.e., below, near, and above the barrier. For this
purpose, the cross-section data were calculated by using the CCFULL code
\cite{37}, taking into account the excitations of 2$^{+}$ and
3$^{-}$ states of the target and projectile nuclei. The parameters
applied to describe the excitations of these low-lying states for
the chosen nuclei were taken from Refs. \cite{38,39}. The results of the
calculations based on the potentials obtained from the different forces
are shown in Fig. 5 in both logarithmic and linear scales. It can be
seen that the theoretical results are obviously influenced by the
incompressibility of the Skyrme forces. The interaction potentials
calculated from the forces with smaller incompressibility
values precisely describe the experimental fusion cross sections
\cite{40} at low energies, but cannot explain the data at
above-barrier energies. Furthermore, it is evident that the potentials obtained from the forces associated with
higher incompressibility values can accurately reproduce the fusion cross sections at high
energies; however, they cannot predict the data at subbarrier energies. To be more precise, based on
this observation, it is found that the Skyrme forces associated with
the nuclear incompressibility values $ \sim$234-248 MeV can reproduce the fusion cross sections of
\textsuperscript{16}O+\textsuperscript{208}Pb at energies below and
near the barrier, the Skyrme force resulting in $K$=270 MeV can
explain the experimental data at energies in the vicinity and nearly above the
barrier, and the forces leading to $K>$300 MeV can be used to
predict the fusion cross sections at energies above the barrier and
at higher energies.

To demonstrate the importance of the density parameters in these
calculations, the fusion cross sections of the chosen system
were also computed by using the potentials derived from the different
forces and the same sets of density parameters, which were obtained
with the SkP Skyrme force \cite{41} for the interacting nuclei. The
calculated fusion cross sections are illustrated in Fig. 6.  As one
can observe, in this case, the experimental and theoretical fusion cross sections are not in agreement, which clearly shows that the density
parameters play a key role in reproducing the experimental fusion
data and in examining the sensitivity of the fusion cross sections to the incompressibility
value at different bombarding energies.

In addition, to study the nuclear matter incompressibility in the
\textsuperscript{16}O+\textsuperscript{208}Pb system, the fusion barrier distribution, $
d^{2}(E\sigma_{fus})/dE^{2} $, for this system was computed. Figure 7 shows the barrier distributions calculated by using the cross sections derived from different Skyrme forces. The theoretical barrier distributions display almost a similar behavior as found in the prediction of the fusion cross sections. The experimental representation of the barrier distribution at high
energies can be better explained by using the cross sections derived from the Skyrme forces yielding higher $K$ values. However, at low energies, the agreement between the experimental and theoretical barrier distributions is achieved by using the data computed from the forces with smaller values for $K$.

According to the results, one can indicate the variation in the nuclear-matter incompressibility within the \textsuperscript{16}O+\textsuperscript{208}Pb system at different energies. To illustrate this, based on the best agreement achieved between the calculated and experimental fusion cross sections at each energy, the predicted values of the nuclear
incompressibility at different bombarding energies are displayed in Fig. 8. As seen, the incompressibility of the nuclear matter
increases by increasing the bombarding energy.

At each energy, the corresponding temperature $T$ of the compound
nucleus, which is displayed on the top horizontal axis of this
figure, was calculated by the following formula \cite{42,43}:

\begin{equation} \label{10}
E^{*}=E_{c.m.}+Q_{in}=\dfrac{1}{a}AT^{2}-T,
\end{equation}
where, $E^{*}$, $E_{c.m.}$, and $Q_{in}$ are the excitation energy
of the compound nucleus, the center-of-mass energy of the projectile
nucleus, and the entrance-channel $Q$ value, respectively. Moreover,
in this equation $ a $ = 9 or 10 for intermediate mass or superheavy
systems.

It can be observed that, by increasing the bombarding energy, the temperature of the compound nucleus increases as well. Therefore, one can expect a variation in the mean-field of the compound system and, consequently, in the property of the nuclear matter as the bombarding energy increases.
\\

\noindent{\bf\small{Efficiency of the described method for other systems}}

By using the suitable Skyrme forces and their corresponding density distributions, the described method can be also applied to study the incompressibility of nuclear matter in other fusion reactions. To show the efficiency of this method for other systems, we briefly discuss the results of the theoretical fusion cross sections for the \textsuperscript{40}Ca+\textsuperscript{90}Zr system.

The potentials derived from the SkT4, SkT1*, SK255, and SK272 Skyrme forces \cite{44,45}, which yield the $K$ values in the range between 235 and 272 MeV and can reasonably describe the properties of the interacting nuclei, were selected as the best choices to describe the fusion cross sections of the system at different bombarding energies. By using these potentials, the theoretical fusion cross sections of the \textsuperscript{40}Ca+\textsuperscript{90}Zr system were computed with the CCFULL code. Figure 9 compares the theoretical results with the experimental data \cite{46}. The agreement between the experimental and theoretical fusion cross sections derived from the forces with different incompressibility values shows that, as the bombarding energy increases, the nuclear matter becomes more incompressible.
\\

\
\noindent{\bf{IV. Conclusions}\\}
\

The present study examined the variation in the incompressibility of nuclear matter in the \textsuperscript{16}O+\textsuperscript{208}Pb
fusion reaction. To this end, the interaction potential of the
system was calculated by using different Skyrme interactions with the $K$ values ranging from 234 to 370 MeV in the energy-density formalism.
Analysis of the potentials indicated that the use of Skyrme forces with higher nuclear incompressibility values results
in greater barrier heights whose corresponding positions are shifted
to closer distances between the interacting nuclei. The fusion cross sections of
the chosen system were computed by using the ion-ion potentials and the CCFULL code. The results revealed that the experimental cross sections at subbarrier energies can be accurately described by the potentials derived
from the forces with smaller $K$ values. On the other hand, the data at
higher energies can be satisfactorily explained by the potentials
obtained from the forces associated with higher K values.
This trend suggests that an exact fit to fusion
cross-section data in different energy ranges can be achieved by using
forces with different incompressibility values. 

Based on the calculations made by the Skyrme energy density formalism and the CCFULL code, one can conclude that nuclear matter during the fusion process changes from less-incompressible matter at low energies to more-incompressible matter at higher energies. In addition, it is worth
mentioning that the applied method enables analysis of the
property of nuclear matter in the fusion process at different
bombarding energies based on a static model.
\\
\\


\newpage
\noindent{\bf {FIGURE CAPTIONS}}\\
\\
Fig. 1. The neutron and proton density distributions of (a) the \textsuperscript{16}O and (b) \textsuperscript{208}Pb nuclei obtained
by using the SkSC4 and SI Skyrme interactions in the HFB approximation.\\
\\
\\
Fig. 2. The percentage relative deviations, i.e., $ \left|(Theo. - Exp.)/Exp.\right|\times100$,
of (a) the theoretical binding energies and (b) root-mean-square charge radii from their experimental
data for the \textsuperscript{16}O and \textsuperscript{208}Pb nuclei. The incompressibility values
corresponding to the Skyrme forces are displayed on the top horizontal axis.\\
\\
\\
Fig. 3. The calculated diffuseness parameters of the neutron and
proton density distributions, $a_{n,p}$, for (a) the
\textsuperscript{16}O and (b) \textsuperscript{208}Pb nuclei.
The incompressibility values corresponding to the Skyrme forces are displayed on the top horizontal axis.\\
\\
\\
Fig. 4. (a) The theoretical fusion barrier heights and (b) positions calculated from
different Skyrme forces for the \textsuperscript{16}O+\textsuperscript{208}Pb system.
The incompressibility values corresponding to the Skyrme forces are displayed on the top horizontal axis.\\
\\
\\
Fig. 5. The fusion cross sections of the \textsuperscript{16}O+\textsuperscript{208}Pb system calculated with 
the potentials obtained from different Skyrme forces. The experimental data were taken from Ref. \cite{40}.\\
\\
\\
Fig. 6. The fusion cross sections of the \textsuperscript{16}O+\textsuperscript{208}Pb system calculated with  
the potentials derived from different Skyrme forces and the density parameters obtained from the SkP Skyrme force.\\
\\
\\
Fig. 7. The fusion barrier distributions for the \textsuperscript{16}O+\textsuperscript{208}Pb system
calculated by using the cross sections derived from different Skyrme forces and their corresponding density distributions.\\
\\
\\
Fig. 8. The predicted values of the nuclear matter incompressibility
in the \textsuperscript{16}O+\textsuperscript{208}Pb system at different
bombarding energies. The temperature of the compound nucleus
corresponding to each energy is displayed on the top horizontal
axis.\\
\\
\\
Fig. 9. The fusion cross sections of the \textsuperscript{40}Ca+\textsuperscript{90}Zr system calculated with
the potentials obtained from the SkT4, SkT1*, SK255, and SK272 Skyrme forces and their corresponding density distributions. The experimental data were taken from Ref. \cite{46}.
\\
\newpage

\begin{figure}
\begin{center}
\includegraphics{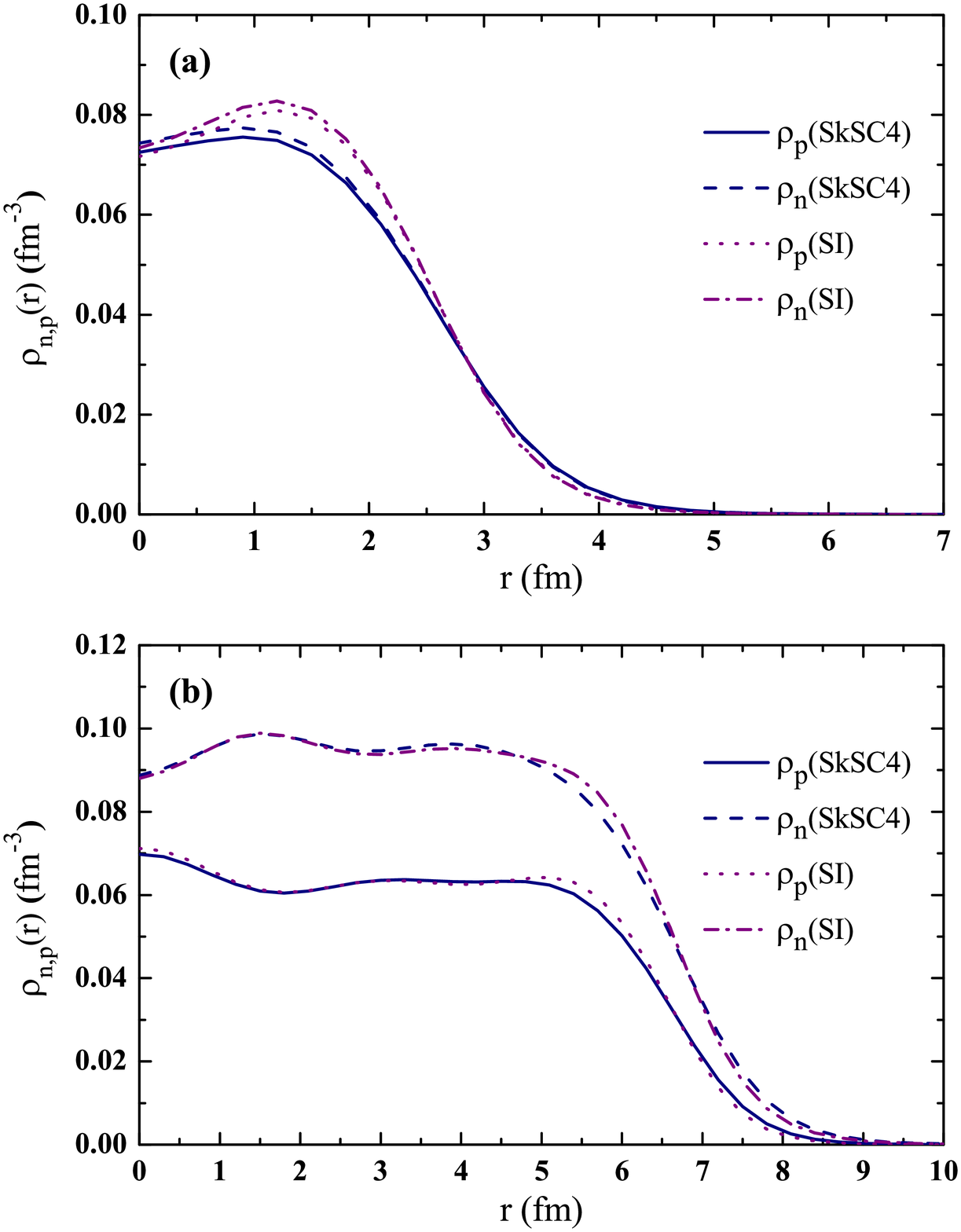}
\end{center}
\vspace{15cm} \caption{}
\end{figure}

\begin{figure}
\begin{center}
\includegraphics{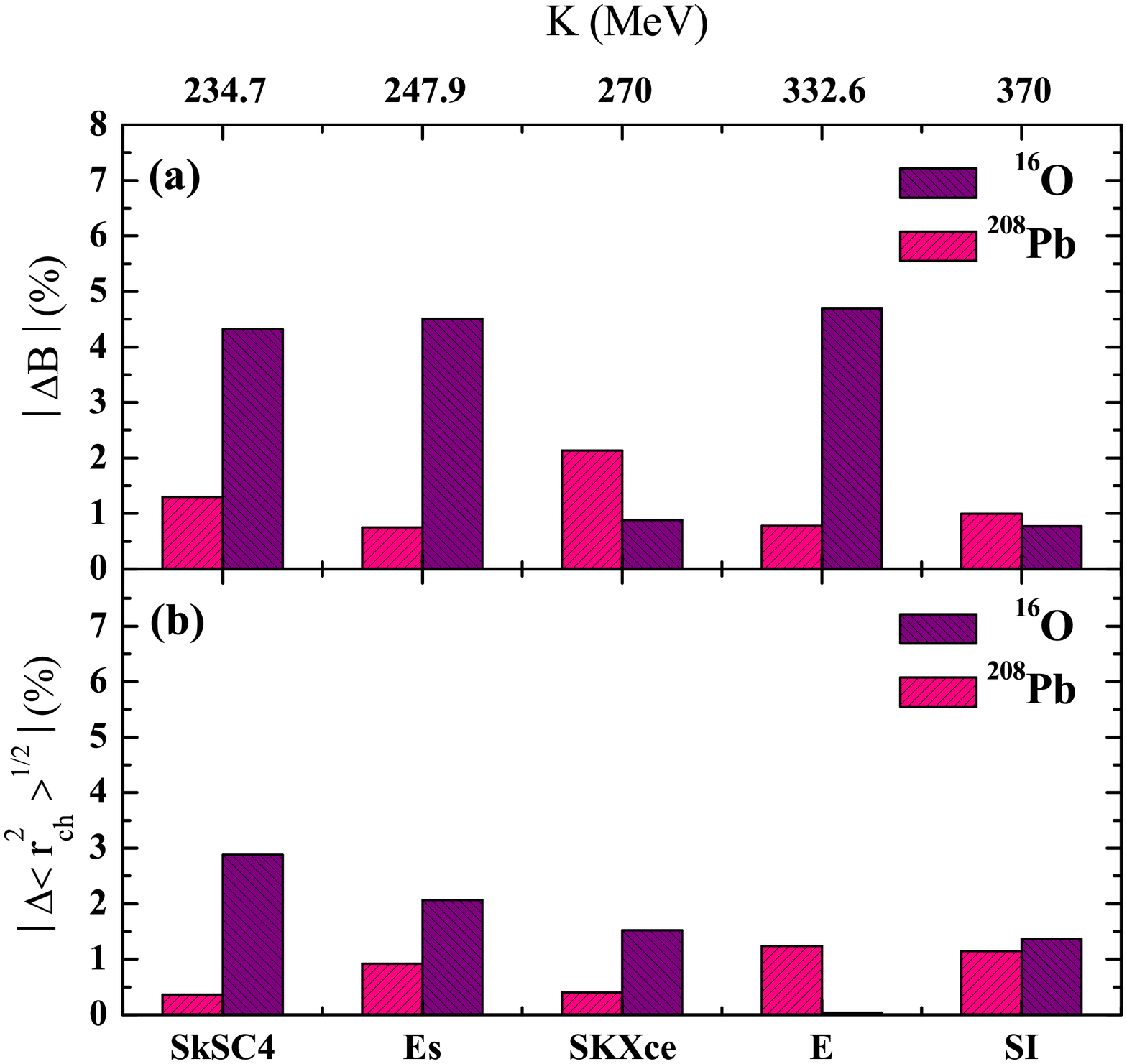}
\end{center}
\vspace{15cm} \caption{}
\end{figure}

\begin{figure}
\begin{center}
\includegraphics{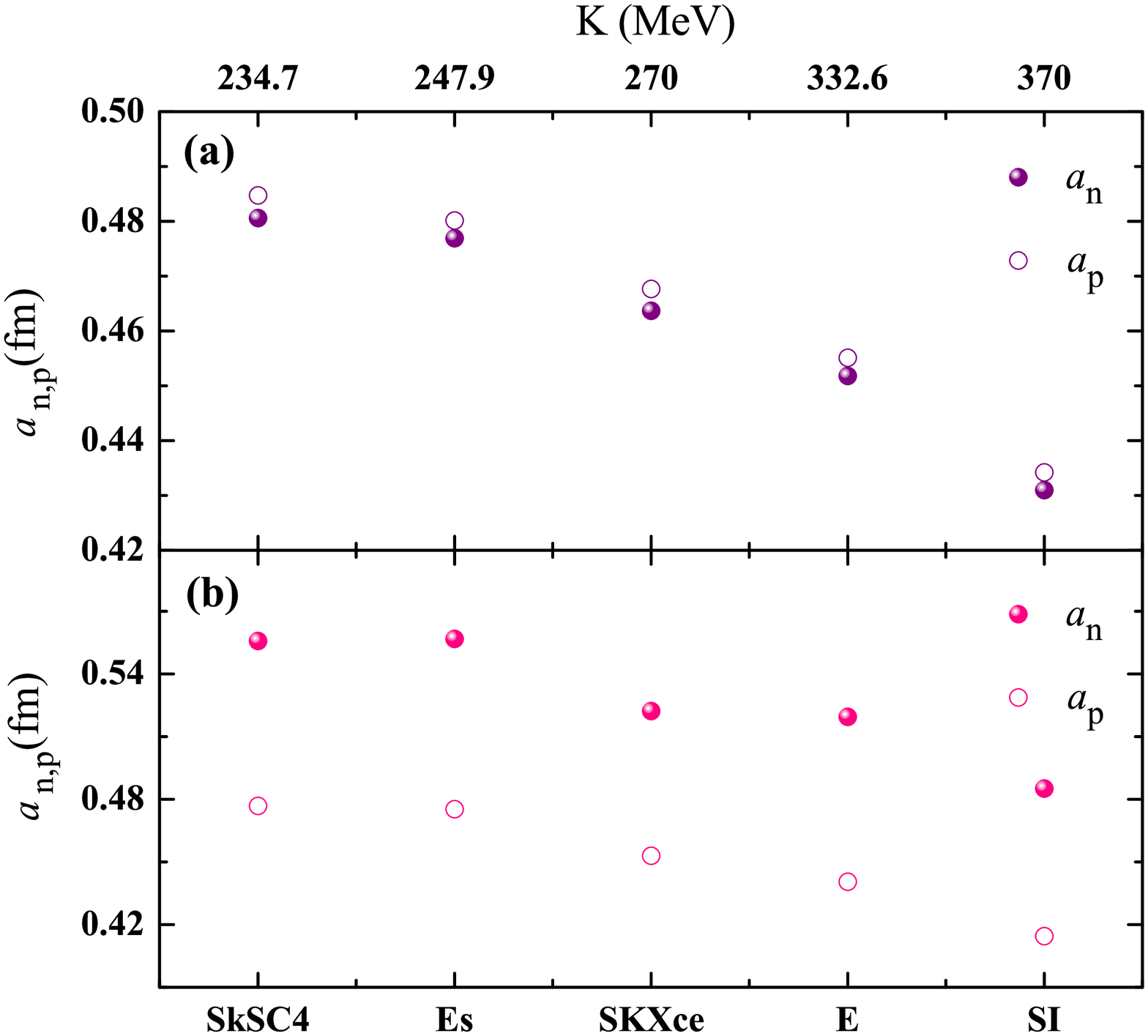}
\end{center}
\vspace{15cm} \caption{}
\end{figure}

\begin{figure}
\begin{center}
\includegraphics{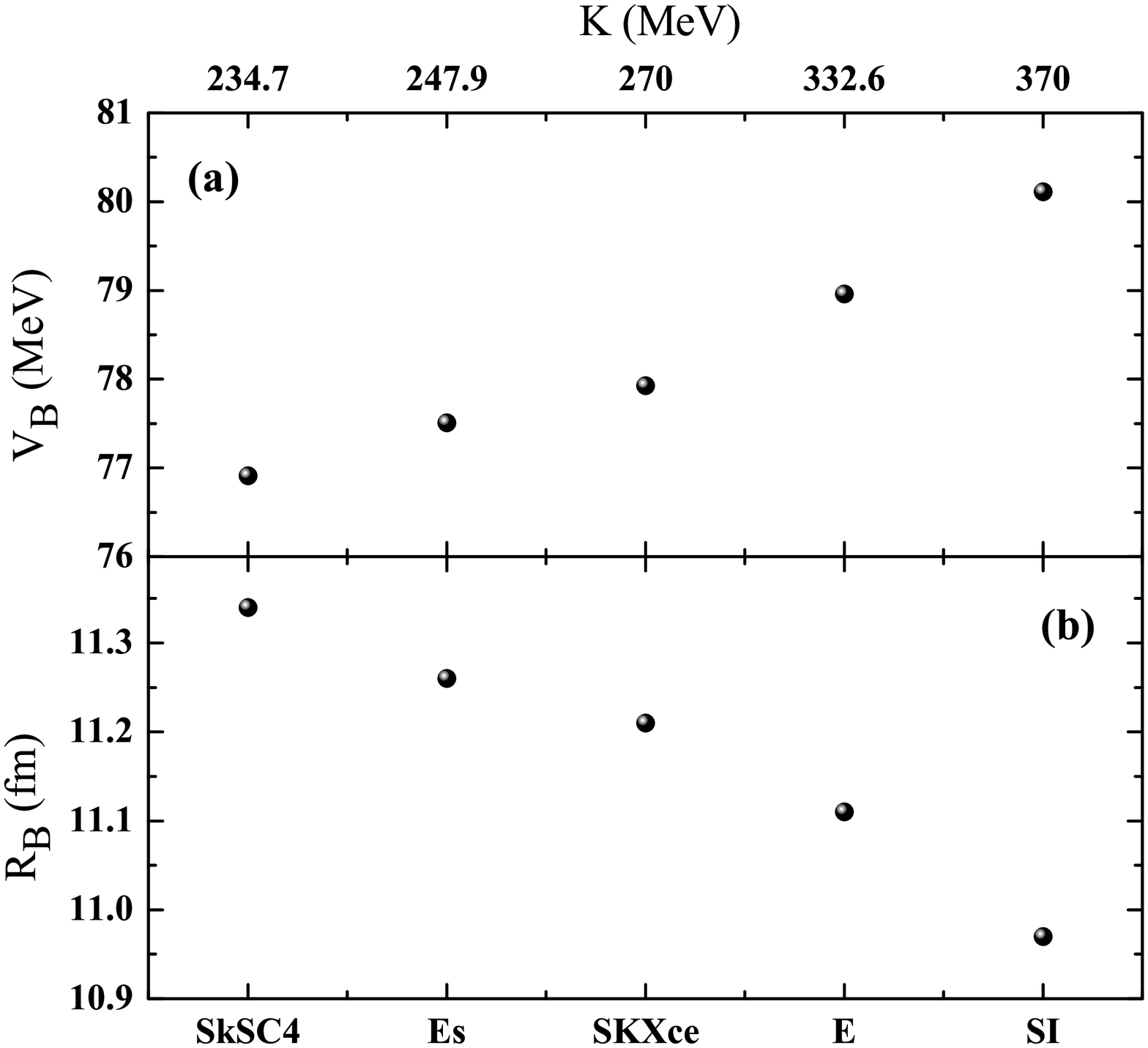}
\end{center}
\vspace{15cm} \caption{}
\end{figure}

\begin{figure}
\begin{center}
\includegraphics{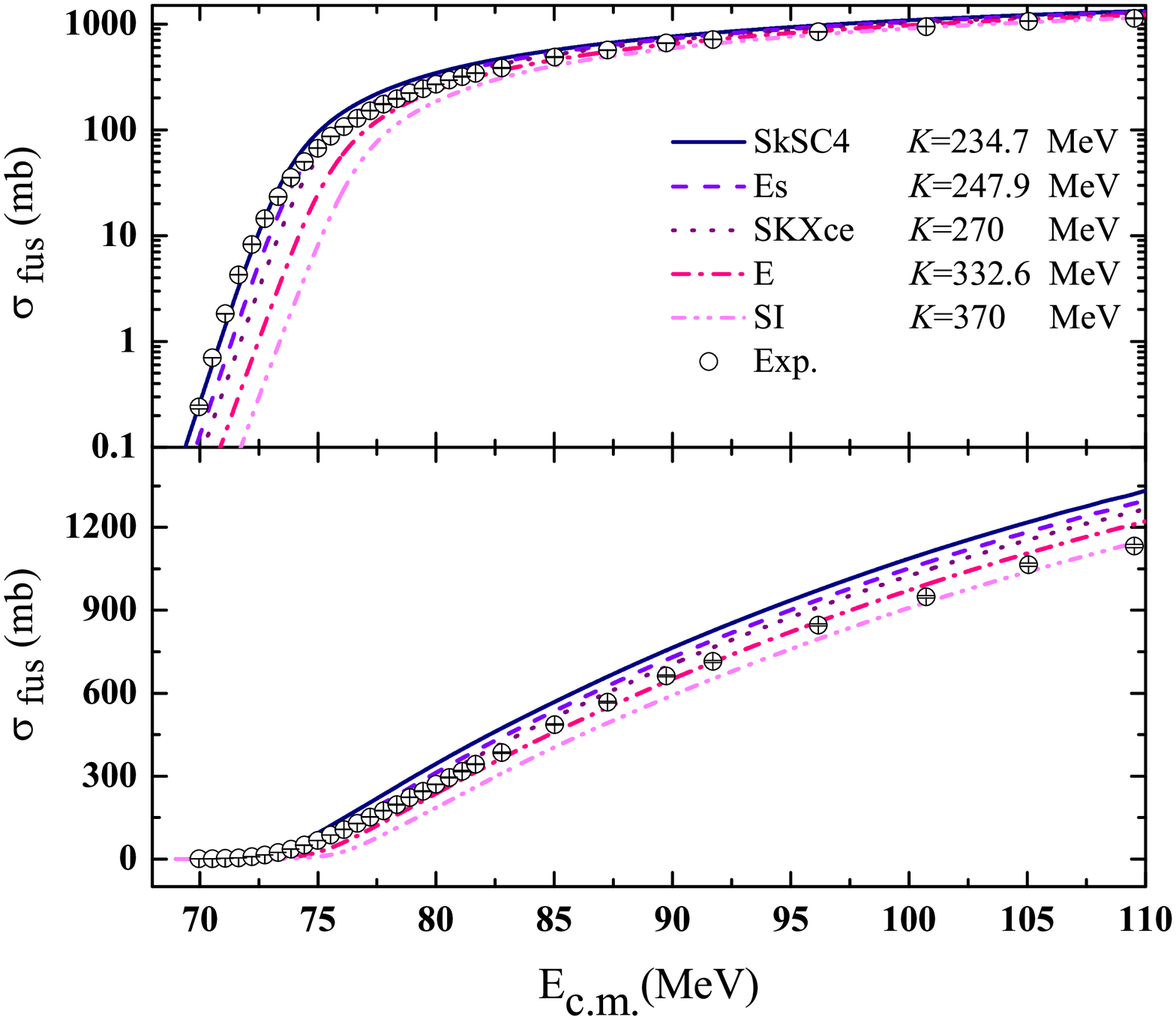}
\end{center}
\vspace{15cm} \caption{}
\end{figure}

\begin{figure}
\begin{center}
\includegraphics{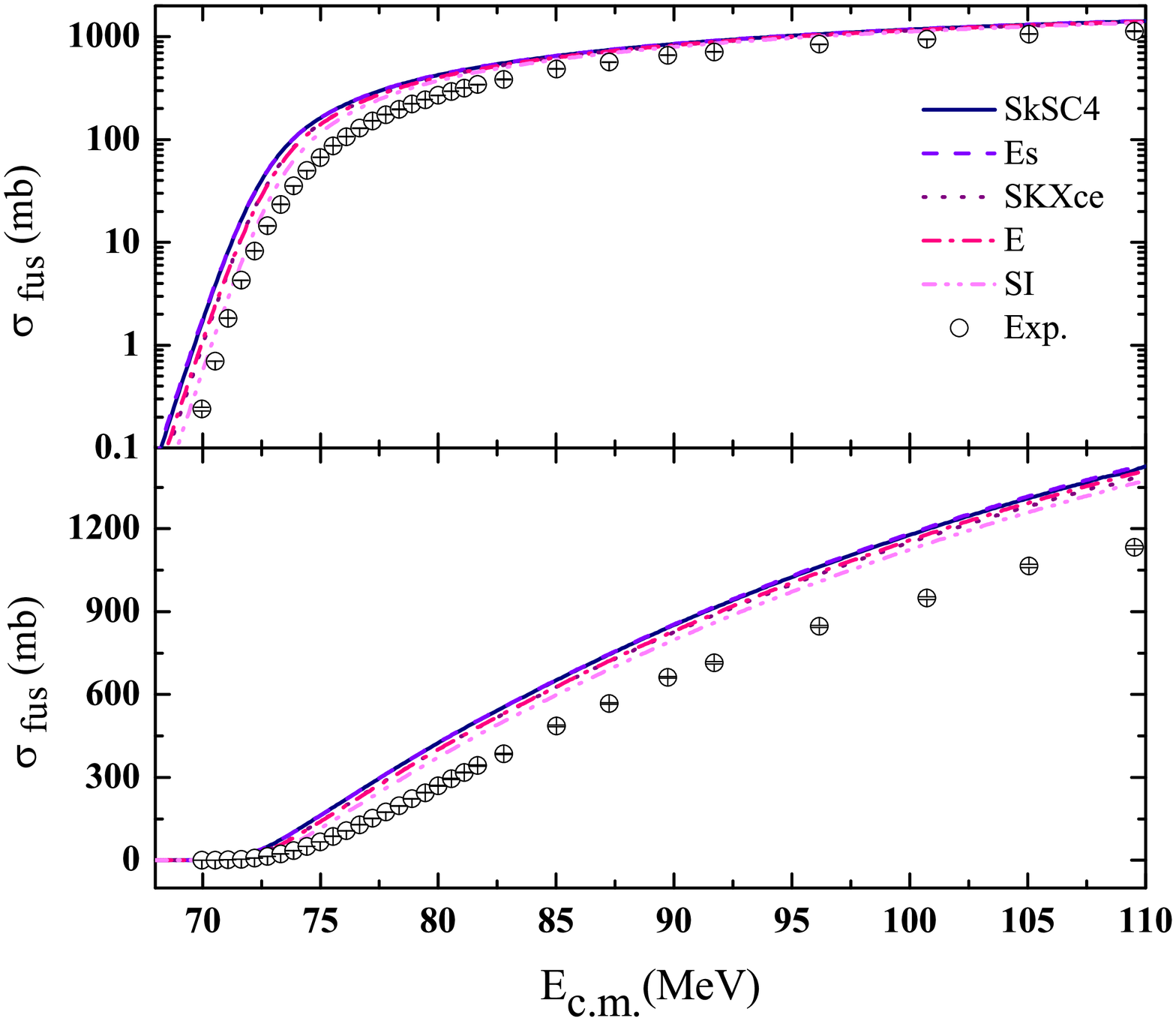}
\end{center}
\vspace{15cm} \caption{}
\end{figure}

\begin{figure}
\begin{center}
\includegraphics{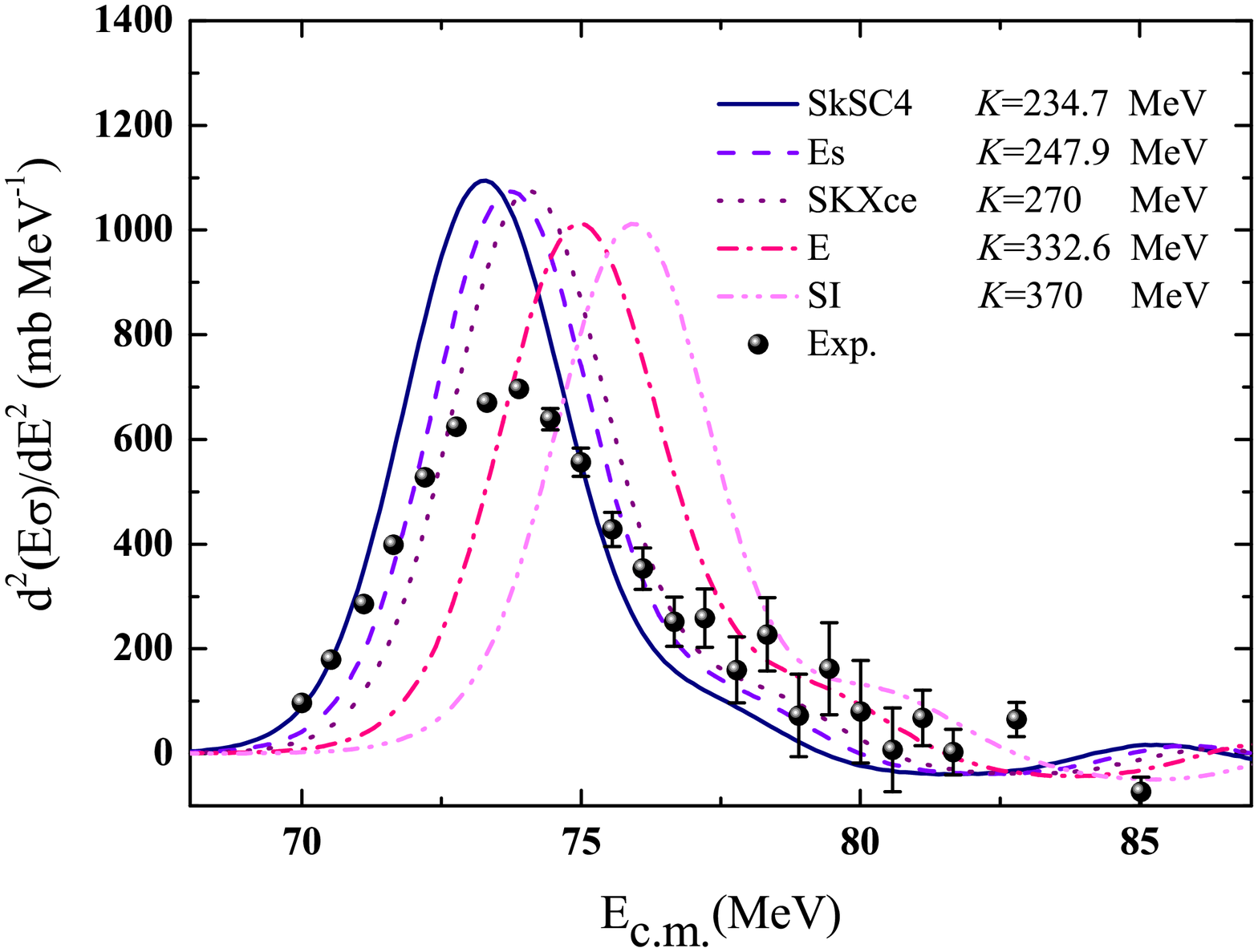}
\end{center}
\vspace{15cm} \caption{}
\end{figure}

\begin{figure}
\begin{center}
\includegraphics{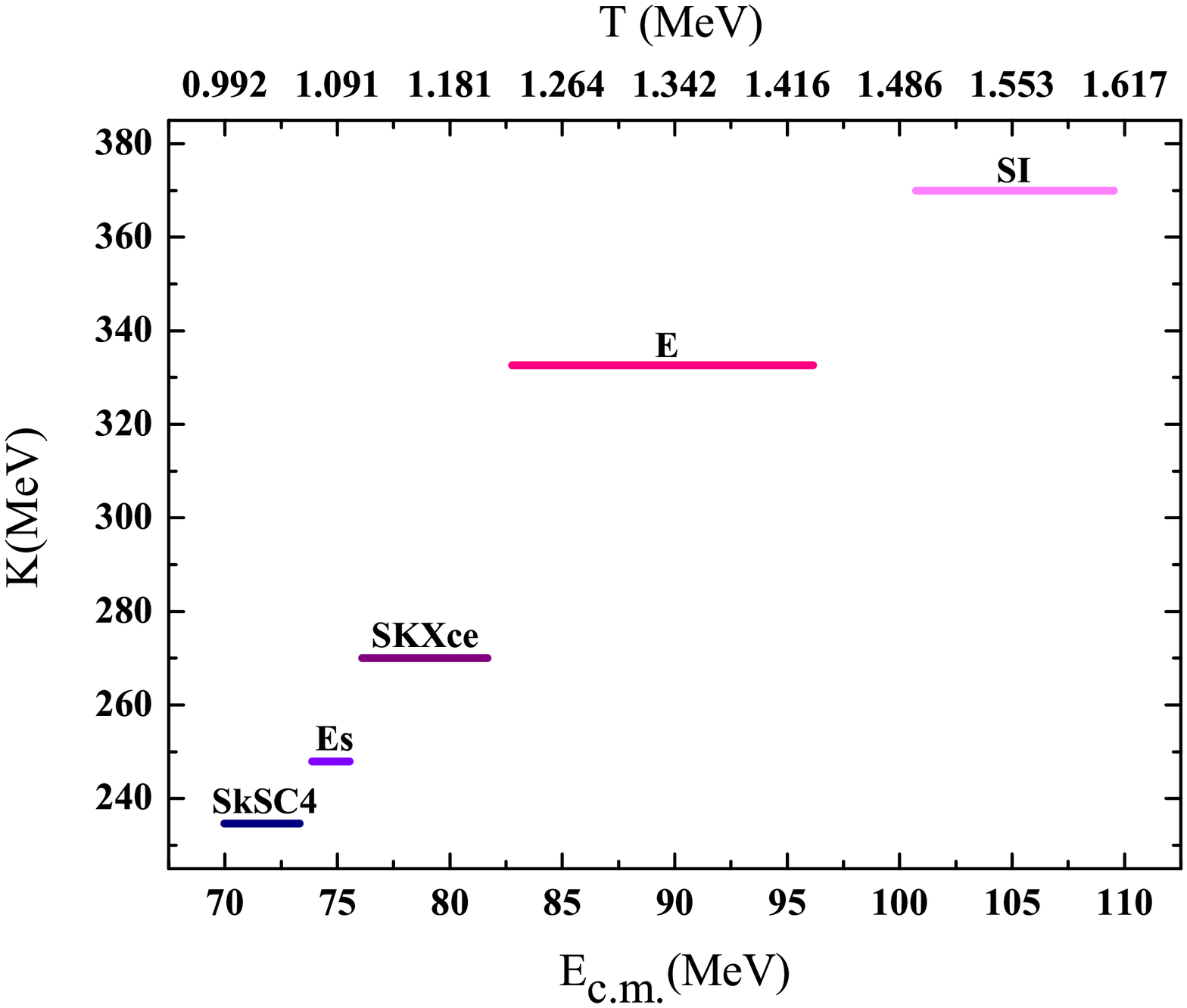}
\end{center}
\vspace{15cm} \caption{}
\end{figure}

\begin{figure}
\begin{center}
\includegraphics{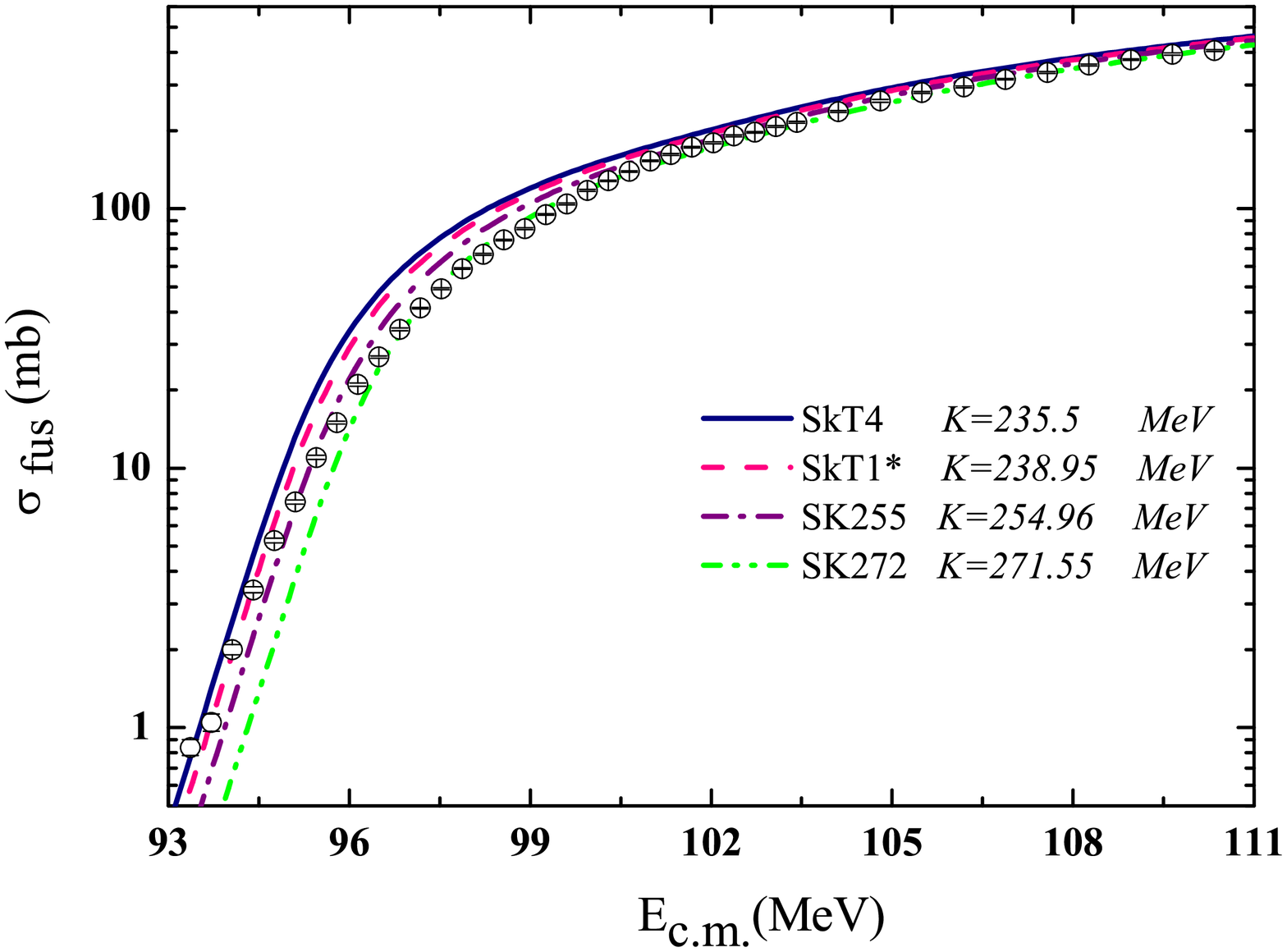}
\end{center}
\vspace{15cm} \caption{}
\end{figure}

\end{document}